%Paper: hep-ph/9401345
%From: martin john savage <savage@thepub.phys.cmu.edu>
%Date: Thu, 27 Jan 94 16:14:38 -0500

\input epsf

\ifx\epsffile\undefined\message{(FIGURES WILL BE IGNORED)}
\def\insertfig#1#2{}% null macro
\else\message{(FIGURES WILL BE INCLUDED)}
\def\insertfig#1#2{{{\baselineskip=4pt
\midinsert\centerline{\epsfxsize=\hsize\epsffile{#2}}{{
\centerline{#1}}}\medskip\endinsert}}}
\fi
\input harvmac
%%%%%%%%%%%%%%%%%%%%%%%%%%%%%%%%%%%%%%%%%%%%%%%%%%%%%%%%%%%%%%%%%%%%%%

%
%  UCSD macros to overwrite some of the definitions in harvmac.tex
%  (include after harvmac.tex)
%  last modified 4/92
%
%%%%%%%%%%%%%%%%%%%%%%%%%%%%%%%%%%%%%%%%%%%%%%%%%%%%%%%%%%%%%%%%%%%%%%
%
%
% modify the output routine for the little format
%
\ifx\answ\bigans
\else
\output={

\almostshipout{\leftline{\vbox{\pagebody\makefootline}}}\advancepageno

}
\fi
%
%
% address
%

%
% grant numbers
%

%
% preprint number
%
\def\UCSD#1#2{\noindent#1\hfill #2%
\bigskip\supereject\global\hsize=\hsbody%
\footline={\hss\tenrm\folio\hss}}% restores pagenumbers
%
% abstract
%
\def\abstract#1{\centerline{\bf Abstract}\nobreak\medskip\nobreak\par
#1}
%
%
% titlefont
%
%
\edef\tfontsize{ scaled\magstep3}
 \tfontsize  \tfontsize
\font\titlermss=cmr5 \tfontsize \font\titlei=cmmi10 \tfontsize
\font\titleis=cmmi7 \tfontsize \font\titleiss=cmmi5 \tfontsize
\font\titlesy=cmsy10 \tfontsize \font\titlesys=cmsy7 \tfontsize
\font\titlesyss=cmsy5 \tfontsize  \tfontsize
\skewchar\titlei='177 \skewchar\titleis='177 \skewchar\titleiss='177
\skewchar\titlesy='60 \skewchar\titlesys='60 \skewchar\titlesyss='60
\scriptscriptfont0=\titlermss
%\textfont1=\titlei \scriptfont1=\titleis
\scriptscriptfont1=\titleiss
%\textfont2=\titlesy \scriptfont2=\titlesys
\scriptscriptfont2=\titlesyss
%\textfont\itfam=\titleit \def\it{\fam\itfam\titleit}\rm}
%
%
% math symbols
%
%---------------------------------------------------------------------

%
\def\inv{^{\raise.15ex\hbox{${\scriptscriptstyle -}$}\kern-.05em 1}}
  %prime
\def\lbar{{\lower.35ex\hbox{$\mathchar'26$}\mkern-10mu\lambda}}
%lambda bar

%
%
% various slashed symbols
%
%
\def\slash#1{\rlap{$#1$}/} % slashes a character
\def\dsl{\,\raise.15ex\hbox{/}\mkern-13.5mu D} %this one can be subscripted
\def\delsl{\raise.15ex\hbox{/}\kern-.57em\partial}
\def\Ksl{\hbox{/\kern-.6000em\rm K}}
\def\Asl{\hbox{/\kern-.6500em \rm A}}
\def\Dsl{\hbox{/\kern-.6000em\rm D}} %roman D
\def\Qsl{\hbox{/\kern-.6000em\rm Q}}
\def\gradsl{\hbox{/\kern-.6500em$\nabla$}}
%
% space and backspace in l mode
%
\def\lspace{\ifx\answ\bigans{}\else\qquad\fi}
\def\lbspace{\ifx\answ\bigans{}\else\hskip-.2in\fi} % $$\lbspace...$$
%
%     boxes an equation
%
\def\boxeqn#1{\vcenter{\vbox{\hrule\hbox{\vrule\kern3pt\vbox{\kern3pt
        \hbox{${\displaystyle #1}$}\kern3pt}\kern3pt\vrule}\hrule}}}
%
%     draw a little box (end of proof symbol)
%     e.g. \mbox{.1}{.1}
%
\def\mbox#1#2{\vcenter{\hrule \hbox{\vrule height#2in
\kern#1in \vrule} \hrule}}
%
%
%
%     curly letters
%
   %curly letters

%
%
%
%     derivatives
%
%

%

\def\bar#1{\overline{#1}}

\def\darr#1{\raise1.5ex\hbox{$\leftrightarrow$}\mkern-16.5mu #1}

%
 %pound sterling
%
 %puts a small half in a displayed eqn
\def\frac#1#2{{\textstyle{#1\over #2}}} %puts a small fraction
%in a displayed eqn
%
%
%     various math operators
%
%

%
%
%
%

%
%       relations
%
\def\ltap{\ \raise.3ex\hbox{$<$\kern-.75em\lower1ex\hbox{$\sim$}}\ }
\def\gtap{\ \raise.3ex\hbox{$>$\kern-.75em\lower1ex\hbox{$\sim$}}\ }
\def\gl{\ \raise.5ex\hbox{$>$}\kern-.8em\lower.5ex\hbox{$<$}\ }
\def\roughly#1{\raise.3ex\hbox{$#1$\kern-.75em\lower1ex\hbox{$\sim$}}}

%
%
%       This defines et al., i.e., e.g., cf., etc.

%

%
\def\np#1#2#3{{Nucl. Phys. } B{#1} (#2) #3}
\def\pl#1#2#3{{Phys. Lett. } {#1}B (#2) #3}
\def\prl#1#2#3{{Phys. Rev. Lett. } {#1} (#2) #3}
\def\physrev#1#2#3{{Phys. Rev. } {#1} (#2) #3}

\relax

\def\hbar{\bar h_Q}

\def\qsl{\hbox{/\kern-.5600em {$q$}}}
\def\ksl{\hbox{/\kern-.5600em {$k$}}}

\def\({\left(}
\def\){\right)}

\def\OMIT#1{}
\def\frac#1#2{{#1\over#2}}

\def\three{$\overline{\bf 3}$}
\def\six{${\bf 6}$}
\hbadness=10000

\noblackbox
\vskip 1.in
\centerline{{\titlefont{Magnetic Moment of the $\Lambda_c$, $\Xi_{c1}^+$ and
$\Xi_{c1}^0$}}}
\medskip
\vskip .5in
\centerline{Martin J.~Savage}
\medskip
{\it{ \centerline{Department of Physics, Carnegie Mellon University, Pittsburgh
PA 15213} }}

\vskip .5in

\abstract{
The magnetic moment of the $\Lambda_c$, $\Xi_{c1}^+$
and $\Xi_{c1}^0$ vanish when the charm quark mass is taken to infinity
because the light degrees of freedom are in a spin zero configuration.
The heavy quark spin-symmetry violating contribution from the
light degrees of freedom starts at order
$1/m_c$, the same order as the contribution from the heavy charm quark.
We compute the leading long-distance contribution to the magnetic moments
from the spin-symmetry breaking
$\Sigma_c^*-\Sigma_c $ mass splitting in chiral perturbation theory.
These are nonanalytic in the pion mass and arise from calculable one-loop
graphs.
Further, the difference between the magnetic moments of the charged charmed
baryons is independent of the charm quark mass and of the subleading local
counterterm.
 }

\vfill
\UCSD{\vbox{
\hbox{CMU-HEP 94-06}
\hbox{DOE-ER/40682-60}}
}{January  1994}
\eject

The strong dynamics of a quark are greatly simplified in the limit that
its mass becomes much greater than the scale of strong interactions.
\ref\iwhq{N. Isgur and M.B.Wise, \pl{232}{1989}{113};
\pl{237}{1990}{527}.}\nref\ehhq{E. Eichten and B. Hill,
\pl{234}{1990}{511}.}-\ref\gehq{H. Georgi, \pl{240}{1990}{447}.} .
Observables involving heavy quarks have a power series
expansion in inverse powers of the quark mass and
perturbative strong interactions.
In the infinite mass limit, baryons containing a single heavy quark can be
classified according to the spin of the light degrees of freedom.
The \three\  of charmed baryons ($\Lambda_c, \Xi^+_{c1}$ and
$\Xi^0_{c1}$) have $s_l=0$ and the spin of the baryon is carried entirely
by the heavy quark.
Since the magnetic moment of the heavy quark vanishes
as its mass becomes infinite so does the magnetic moment of these baryons.
In chiral perturbation theory this translates into there being no magnetic
moment counterterm
from the electromagnetic current of the light quarks that preserves heavy quark
spin symmetry.
In this work we point out that the leading contribution to the
magnetic moment of the \three\ charmed baryons from long-distance
physics is calculable in chiral perturbation theory.
It arises from the spin-symmetry breaking
$\Sigma_c^*-\Sigma_c$ mass splitting and has nonanalytic dependence on
the mass of the pion.

Heavy quark symmetry and chiral symmetry are combined together
in order to describe the soft hadronic interactions of hadrons containing a
heavy quark
\ref\mbwchir{M. B. Wise, \physrev{D45}{1992}{2188}.}\nref\burd{G. Burdman and
J. Donoghue, \pl{208}{1992}{287}.}\nref\yan{T. M. Yan etal.,
\physrev{D46}{1992}{1148}.}-\ref\cho{P. Cho, \pl{285}{1992}{145}.} .
We are only concerned with dynamics of heavy baryons and so we will not
discuss the lagrangian for heavy mesons, we refer the reader to
\ref\mbwlouise{M.B. Wise, Lectures given at the CCAST Symposium on Particle
Physics at the Fermi Scale, May 1993.}
for a review.
Light degrees of freedom in the ground state of a baryon containing one heavy
quark
can have $s_l=0$ corresponding to a member of the
flavour $SU(3)$ \three , $T_i (v)$
or they can have $s_l=1$ corresponding to a member of the flavour $SU(3)$ \six\
 ,
$S^{ij}_\mu (v)$.
In the latter case, the spin of the light degrees of freedom can be combined
with the
spin of the heavy
quark to form both $J=3/2$ and $J=1/2$ baryons, which are degenerate in the
$m_Q\rightarrow \infty$ limit.
Using the notation of \cho\ we define the fields,
\eqn\tripsex{ \eqalign{
S^{ij}_\mu (v) = & {1\over\sqrt{3}} (\gamma_\mu + v_\mu) \gamma_5
{1\over 2} (1+\slash{v}) B^{ij}  \ + \ {1\over 2}(1+\slash{v}) B^{*ij}_\mu \cr
T_i (v) = & {1\over 2}(1+\slash{v}) B_i } \ \ \ \ ,}
where the $J=1/2$ charmed baryons of the \six\ are assigned to
the symmetric tensor $B^{ij}$
\eqn\sex{\eqalign{
B^{11} = \Sigma_c^{++}\ ,\ B^{12} = {1\over\sqrt{2}}\Sigma_c^+\ ,\ B^{22} =
\Sigma_c^0 \ ,\cr
B^{13} = {1\over\sqrt{2}}\Xi_{c2}^+ \ ,\ B^{23} = {1\over\sqrt{2}}\Xi_{c2}^0 \
,\
B^{33} = \Omega_c^0 }\ \ \ .}
The $J=3/2$ partners of these baryons have the same
$SU(3)_V$ assignment in $B^{*ij}_\mu$.
The charmed baryons of the \three\ representation are assigned to $B_i$ as
\eqn\trip{B_1 = \Xi_{c1}^0 \  , \ B_2 = -\Xi_{c1}^+ \ , \ B_3 = \Lambda_c^+ \ \
\ .}

The chiral lagrangian describing the soft hadronic interaction of these baryons
is given by \cho\
\eqn\lag{ \eqalign{
{\cal L}_Q = &\ \  i\overline{T^i} v\cdot D T_i \ -\  i\overline{S}^\mu_{ij}
v\cdot D S_\mu^{ij}
\ +\ \Delta_0 \overline{S}^\mu_{ij} S_\mu^{ij}\cr
&\ + \ g_3\left(  \epsilon_{ijk}\overline{T}^i (A^\mu)^j_l S^{kl}_\mu + h.c.
\right)
+  \ i g_2\ \epsilon_{\mu\nu\rho\sigma}\overline{S}^\mu_{ik} v^\nu(A^\rho)^i_j
S^{\sigma jk} + ............}  \ \ ,}
where the dots denote operators with more insertions of the light quark mass
matrix,
more derivatives or higher order in the $1/m_Q$ expansion and
$D^\alpha$ is the chiral covariant derivative.
The axial chiral field
$A^\mu = {i \over 2}\left( \xi\partial^\mu\xi^\dagger  -
\xi^\dagger\partial^\mu\xi \right) $
is defined in terms of
$\xi = {\rm exp}\left(iM/f_\pi\right)$
where $M$ is the octet of pseudogoldstone bosons
\eqn\mesons{ M = \left(
\matrix{ {1\over\sqrt{6}}\eta + {1\over\sqrt{2}}\pi^0 & \pi^+ & K^+ \cr
                                                      \pi^- &
{1\over\sqrt{6}}\eta - {1\over\sqrt{2}}\pi^0 &
K^0 \cr
                                                        K^- & \overline{K}^0 &
-{2\over\sqrt{6}}\eta \cr } \right) \ \ \ \ \ ,}
and $f_\pi = 135{\rm MeV}$ is the pion decay constant.
Coupling of the pseudo-Goldstone bosons to the \three\  baryons
is forbidden at lowest order in $1/m_Q$.
Even in the infinite mass limit the $\Sigma_Q^{(*)}$ baryons are not degenerate
with the $\Lambda_Q$ baryons as the light degrees of freedom are
in a different configuration giving rise to an intrinsic mass difference
$\Delta_0$.

The magnetic moment interactions of a heavy quark are described by the lagrange
density
\eqn\heavymag{ {\cal L}_{\rm mag} = {e{\cal Q}\over 4m_Q}
\overline{h}^{(Q)}_v\sigma_{\mu\nu} h^{(Q)}_v F^{\mu\nu}\ \ +\ .........\ \ \
,}
where ${\cal Q}$ is the charge of the heavy quark, $F^{\mu\nu}$ is the
electromagnetic field tensor
and the dots denote terms higher order in $1/m_Q$
\ref\luke{M. Luke, \pl{252}{1990}{447}.}.
Its inclusion into the heavy baryon chiral lagrangian has been discussed
previously
\ref\chogeorgi{P. Cho and H. Georgi,
\pl{296}{1992}{408};\pl{300}{1993}{410}(E).}
\ref\cheng{H.Y. Cheng etal., CLNS-93-1192, hep-ph/9308283 (1993).}.
There is no spin symmetry conserving local counterterm describing
the magnetic moment interactions
of the light degrees of freedom in an $s_l=0$ configuration.
Consequently, at lowest order in the chiral expansion the magnetic
moment of all three \three\ charmed baryons vanish.
However, a magnetic moment for these baryons occurs at order
$1/m_c$ from both the light degrees of freedom and from the charm quark itself.

We write the magnetic moment of the \three\ charmed baryons as
\eqn\mag{ \mu^i = \mu_c + \mu^i_l \ \ \ ,}
where $\mu_c$ is the contribution from the charm quark
and $\mu_l^i$ is the contribution from the light degrees of freedom.
$\mu_c$ is found from forward matrix elements of  \heavymag\
between baryon states and is known by heavy quark spin symmetry.
It is reproduced in the heavy baryon lagrangian by
\eqn\magthree{ {\cal L}^{\overline{\bf 3}}_{\rm heavy} = {e{\cal Q}_c\over
4m_c}
\overline{T}^i\sigma^{\mu\nu}T_i F_{\mu\nu}\ \ \ ,}
where ${\cal Q}_c$ is the charge of the charm quark.
The spin-symmetry breaking $\Sigma_c^*-\Sigma_c$ mass splitting
\foot{This mass splitting is responsible for the leading long-distance
corrections to
$\Lambda_b\rightarrow\Lambda_ce^-\overline{\nu}_e$ at zero-recoil arising from
spin-symmetry breaking
\ref\mjs{M.J. Savage, CMU-HEP 94-04, hep-ph/9401273 (1994).}.
The analogous corrections in the meson sector have been computed in
\ref\randwise{L. Randall and M.B.Wise, \pl{303}{1993}{139}.}
\ref\chowwise{C. Chow and M.B.Wise, \physrev{D48}{1993}{5202}.}.}
gives rise to the formally leading contribution to $\mu_l^i$
through the graph shown in
\fig\magmom{Graphs generating the leading long-distance contribution to the
magnetic moment of \three\ baryons.  The wiggly line denotes a photon,
the dashed line a $\pi$,  and $\Sigma_c^{(*)}$ denotes both $\Sigma_c$
and $\Sigma_c^*$ baryons that can be in the
intermediate state.} .
The formally subleading counterterm has the form
\eqn\counter{{\cal L}_{\rm c.t.}^{\overline{\rm 3}} = {e\beta\over 4 m_c}
\overline{T}^i\sigma^{\mu\nu} Q_i^j T_j F_{\mu\nu}\ \ \ \ \ ,}
where $Q$ is the light quark electromagnetic charge matrix
\eqn\charge{ Q={1\over 3}\left( \matrix{ 2&0&0\cr 0&-1&0\cr 0&0&-1}\right) \ \
\ ,}
and $\beta$ is an unknown constant.
We find that, to this order
\eqn\maglightaa{\eqalign{
\mu_l^3 = & -{1\over 3}{\beta\over m_c}\ \phantom{-  ( \Delta m )\ {2\over
3}{g_3^2\over 16\pi^2 f_\pi^2}
\ J (m_\pi, \Delta_0 )} \ \ \ \ : \Lambda_c \cr
\mu_l^2 = & -{1\over 3}{\beta\over m_c}
                -  ( \Delta m )\ {2\over 3}{g_3^2\over 16\pi^2 f_\pi^2}\ J
(m_\pi, \Delta_0 )
\ \ \ \ : \Xi^+_{c1} \cr
\mu_l^1 = & \phantom{-}\ {2\over 3}{\beta\over m_c}
                +  ( \Delta m )\ {2\over 3}{g_3^2\over 16\pi^2 f_\pi^2}\ J
(m_\pi, \Delta_0 )
\ \ \ \ : \Xi^0_{c1} \cr }\ \ \ .}
The function $J( m,\Delta)$ arising  from a taylor expansion of the loop
integral
in powers of $\Delta m$ (the $\Sigma_c^*-\Sigma_c$ mass splitting ) is given by
\eqn\intint{ J( m,\Delta) = \log\left({m^2\over\Lambda_\chi^2}\right) -
{\Delta\over\sqrt{\Delta^2-m^2}}\log\left( {\Delta - \sqrt{\Delta^2-m^2 +
i\epsilon}\over
\Delta + \sqrt{\Delta^2-m^2+ i\epsilon}}\right) \ \ \ ,}
where we have chosen to evaluate the graphs at the chiral symmetry breaking
scale $\Lambda_\chi$.
We have not shown the contribution to each magnetic moment from loops
involving kaons, as these are suppressed compared to the contribution from pion
loops.
When these terms are included the scale dependence of the total
one-loop contribution is compensated by the scale dependence of $\beta$.
We have neglected any SU(3) breaking in the baryon masses
and assumed that the $\Sigma_c^*-\Sigma_c$ mass splitting $\Delta m$
is equal to the $\Xi_{c2}^*-\Xi_{c2}$ mass splitting.
We can see from \intint\ that when $\Delta_0=0$ the loop contribution has a
true infrared divergence regulated by the pion mass.  However, when
$\Delta_0\ne 0$
the graph is no longer infrared divergent as $m_\pi\rightarrow 0$ and is
regulated by the
intrinsic mass splitting.
The physical spectrum of charmed baryons does not correspond to either
regime and so we keep the full functional dependence of  \intint\ .
This is formally the dominant contribution to the amplitude.

The $\Sigma_c^*$ has not been observed yet
\foot{The SKAT bubble chamber group report a signal for the $\Sigma_c^{*++}$
with a mass of $m_{\Sigma_c^*} = 2530\pm 5 $MeV
\ref\SKAT{V.V. Ammosov etal., contributed paper to the International
Symposium on Lepton and Photon Interactions, Ithaca, 1993.}.
This needs independent verification.}
and in order to get an estimate of the long-distance contribution
we use a  nonrelativistic quark model calculation of the $\Sigma_c^*$ mass
$m_{\Sigma_c^*} = 2494\pm 16$MeV
\ref\paul{R. E. Cutkosky and P. Geiger,\physrev{D48}{1993}{1315}.}.
This value, combined with the experimental measurements of the other relevant
masses
\ref\pdg{Particle Data Group, \physrev{D45}{1992}{1}.}
gives $\Delta m = 41\pm 16 {\rm MeV}$
\foot{For other estimates of $\Delta m$ see
\ref\KROQ{W. Kwong, J. Rosner and C. Quigg, Ann. Rev. Nucl. Sci., 37 (1987)
325.}
\ref\fp{A.F. Falk and M.E. Peskin, SLAC-PUB-6311 (1993).}
and for the present experimental situation see
\ref\argus{H. Albrecht etal., (ARGUS), \pl{211}{1988}{489}.}\nref\cleo{T.
Bowcock etal.,
(CLEO), \prl{62}{1989}{1240}.}-\ref\tpc{J.C. Anjos etal., (Tagged Photon
Collaboration),
\prl{62}{1989}{1721}.}\SKAT .} .
We use this to estimate an intrinsic mass splitting of
$\Delta_0=194\pm 11 {\rm MeV}$ where
the uncertainty depends entirely on that of $m_{\Sigma_c^*}$.
The axial coupling constant $g_3$ is, as yet,  undetermined.
However, in the large-$N_c$  limit of QCD ( $N_c$ being the number of colours)
it has been shown to be related to the $\pi$-N axial coupling constant
$g_3=\sqrt{3\over 2}g_A$ where $g_A=1.25$
\ref\gural{Z. Guralnik, M. Luke and A.V. Manohar, \np{390}{1993}{474}.}
\ref\jenkN{E. Jenkins, \pl{315}{1993}{431}.} .
It is this value of $g_3$ we will use in our estimate.
We find that for $m_c=1700$ MeV and $\beta=0$
the magnetic moments of the \three\  charmed baryons are
\eqn\magmag{ \eqalign{
\mu (\Lambda_c ) \sim  &\ 0.37 \ {\rm N.M.} \cr
\mu (\Xi_{c1}^+) \sim &\ 0.42  \ {\rm N.M.}\cr
\mu (\Xi_{c1}^0) \sim &\ 0.32  \ {\rm N.M.}\cr}
\ \ \ .}
We should point out that while the charm quark mass is not well known
\ref\ls{M. Luke and M.J. Savage, UCSD/PTH 93-25 (1993).}
and this leads to a large uncertainty in the magnetic moments,
the difference between any two of the three magnetic
moments does not depend explicitly on  $m_c$ at this order.
Assuming that $\beta$ can be neglected, these differences
depend only on quantities that can be determined experimentally,
the $\Sigma_c^*-\Sigma_c$ mass splitting and axial coupling constant  $g_3$.
However, it is possible that $\beta$ should not be neglected as the
$\Sigma_c^*-\Sigma_c$ mass splitting is small,
much smaller than the corresponding
splitting in the meson sector and because the intrinsic
$\Sigma_c^{(*)}-\Lambda_c$ mass splitting
suppresses the infrared divergence arising in the chiral limit.
We see from \mag\ and \maglightaa\ that the difference between the magnetic
moment
of the $\Lambda_c$ and the $\Xi_{c1}^+$ depends only on $\Delta m$ and $g_3$,
\eqn\diff{ \mu ({\Xi_{c1}^+}) - \mu ({\Lambda_c}) \sim 0.05 \left({\Delta
m\over 41 {\rm MeV}}\right)
\sqrt{2\over 3}\left({g_3\over g_A}\right) \ \ \ {\rm N.M.}\ \ \ .}
Further, the contribution from both the local counterterm
and loop graph cancel in the
sum of the baryon magnetic moments giving the charm quark magnetic moment
as the average over the baryon moments,
\eqn\sum{\mu_c = {1\over 3}\left( \mu (\Lambda_c ) + \mu (\Xi_{c1}^+) +
 \mu (\Xi_{c1}^0) \right) \ \ \ .}

In conclusion, we have shown that the leading contribution from the
light degrees of freedom to the magnetic moments of the
$\Lambda_c, \Xi_{c1}^+$ and $\Xi_{c1}^0$ baryons
results from the spin symmetry breaking $\Sigma_c^*-\Sigma_c$ mass
splitting and is calculable in chiral perturbation theory.
It is found to depend on the third component of isospin,
vanishing for the $\Lambda_c$ but giving an equal and opposite contribution to
the $\Xi_{c1}^+$ and $\Xi_{c1}^0$.
While suppressed by about a factor of two by the intrinsic $\Sigma_c-\Lambda_c$
mass splitting, the contributions are at the $10\%$ level, depending on the
$\Sigma_c^*-\Sigma_c$ mass splitting and axial coupling constant $g_3$.
There is also a contribution from a formally subleading local counterterm which
may be important
due to the small $\Sigma_c^*-\Sigma_c$ splitting and finite intrinsic
$\Sigma_c-\Lambda_c$ splitting.
Mixing between the baryons of the \six\  and \three\ is not only SU(3)
violating but also vanishes in the heavy quark limit.
Therefore we estimate that the effect of such mixing is small compared to the
terms found in this work.

We have shown that the difference between the charged charmed
baryon magnetic moments is independent of the subleading
counterterm and does not depend explicitly on the charm quark mass.
It is given in terms of experimentally observable quantities, the
$\Sigma_c^*-\Sigma_c$ mass splitting and an axial coupling constant $g_3$.
It is possible that the magnetic moment of the charged charmed baryons will be
measured by the spin precession  that occurs during channeling
in bent crystals
\ref\xtal{D. Chen etal., (E761 Collaboration ), \pl{69}{1992}{3286}.}.
This would allow the difference between the charged
charmed baryon magnetic moments to be measured and compared with this work.
If, in addition, the magnetic moment of the neutral charmed baryon were to be
measured
the magnetic moment of the
charm quark itself could be extracted as it is equal to the average of the
baryon moments
\foot{This is also true for the radiative $D^*\rightarrow D\gamma$ transitions
\ref\jim{J.F. Amundson etal., \pl{296}{1992}{415}.}   }.

The same analysis applies to the magnetic moments of $b$-baryons
in the \three\ of SU(3).
In this case it is the $\Sigma_b^*-\Sigma_b$ mass splitting that gives
rise to the leading
long-distance corrections and the contribution of the $b$ quark is
$\mu_b = -{1\over 3}{1\over m_b}$.

\bigskip

I would like to thank M. Luke and  M.B.Wise
for useful discussions.  This research was supported in part by
the Department of Energy under contract DE--FG02--91ER40682.

\listrefs
\listfigs
\vfill\eject

\insertfig{Figure 1 }{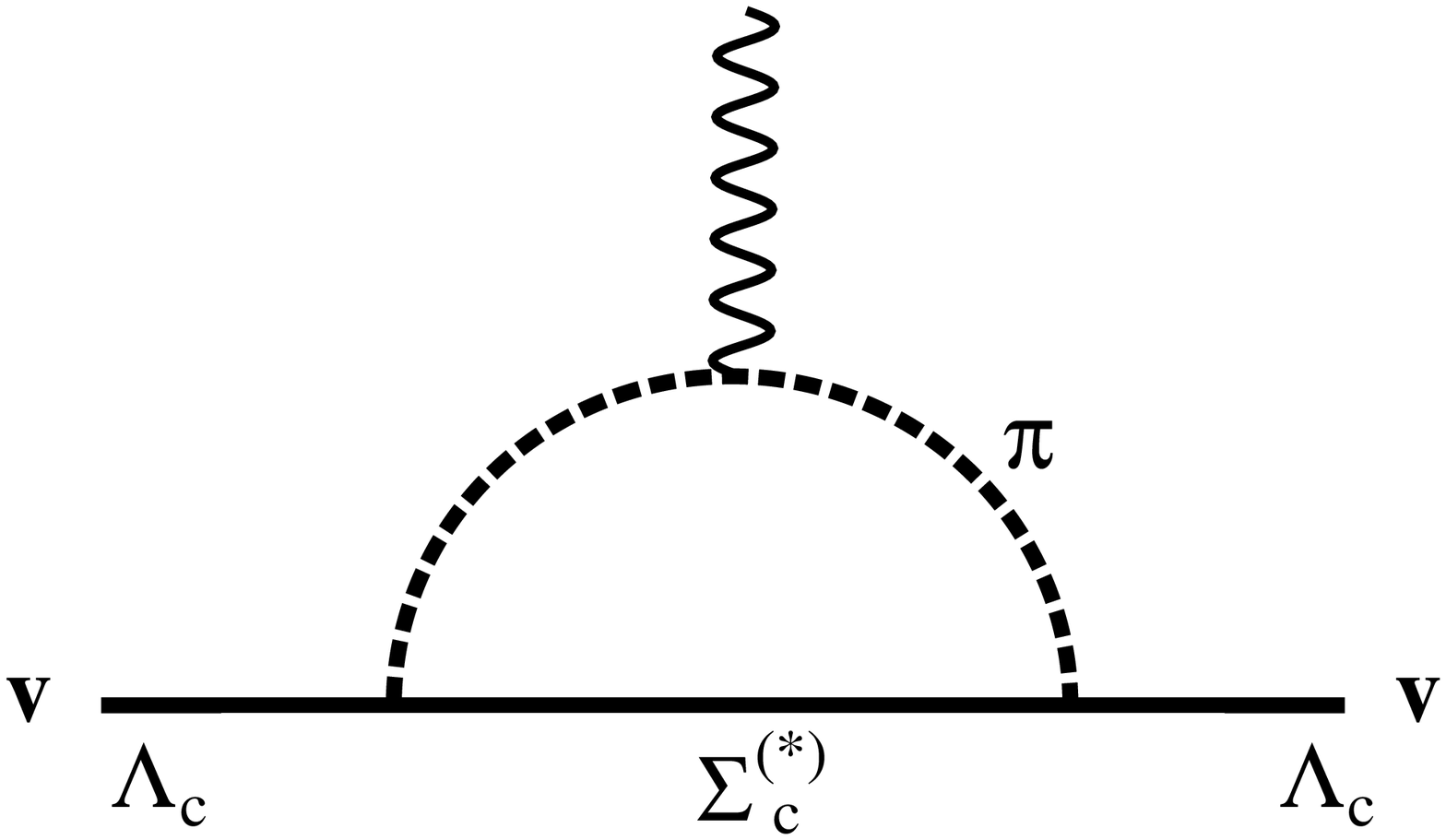}

\vfill\eject
\bye